\documentclass[aps,prl,superscriptaddress,showpacs,preprintnumbers,amsmath,amssymb]{revtex4}

\usepackage{boites}

\usepackage{graphicx} 
\usepackage{dcolumn}  
\usepackage{color}
\graphicspath{{ps}}

\begin{document}


\newcommand{\red}{\textcolor{red}}
\newcommand{\blue}{\textcolor{blue}}
\newcommand{\black}{\textcolor{black}}


\title{Measurement of branching fractions for $B \to J/\psi \eta K$ decays
and search for a narrow resonance in the $J/\psi \eta$ final state}

\noaffiliation
\affiliation{University of the Basque Country UPV/EHU, 48080 Bilbao}
\affiliation{Beihang University, Beijing 100191}
\affiliation{University of Bonn, 53115 Bonn}
\affiliation{Budker Institute of Nuclear Physics SB RAS and Novosibirsk State University, Novosibirsk 630090}
\affiliation{Faculty of Mathematics and Physics, Charles University, 121 16 Prague}
\affiliation{University of Cincinnati, Cincinnati, Ohio 45221}
\affiliation{Deutsches Elektronen--Synchrotron, 22607 Hamburg}
\affiliation{Department of Physics, Fu Jen Catholic University, Taipei 24205}
\affiliation{Justus-Liebig-Universit\"at Gie\ss{}en, 35392 Gie\ss{}en}
\affiliation{II. Physikalisches Institut, Georg-August-Universit\"at G\"ottingen, 37073 G\"ottingen}
\affiliation{Gyeongsang National University, Chinju 660-701}
\affiliation{Hanyang University, Seoul 133-791}
\affiliation{University of Hawaii, Honolulu, Hawaii 96822}
\affiliation{High Energy Accelerator Research Organization (KEK), Tsukuba 305-0801}
\affiliation{Hiroshima Institute of Technology, Hiroshima 731-5193}
\affiliation{Ikerbasque, 48011 Bilbao}
\affiliation{Indian Institute of Technology Guwahati, Assam 781039}
\affiliation{Indian Institute of Technology Madras, Chennai 600036}
\affiliation{Indiana University, Bloomington, Indiana 47408}
\affiliation{Institute of High Energy Physics, Chinese Academy of Sciences, Beijing 100049}
\affiliation{Institute of High Energy Physics, Vienna 1050}
\affiliation{Institute for High Energy Physics, Protvino 142281}
\affiliation{INFN - Sezione di Torino, 10125 Torino}
\affiliation{Institute for Theoretical and Experimental Physics, Moscow 117218}
\affiliation{J. Stefan Institute, 1000 Ljubljana}
\affiliation{Kanagawa University, Yokohama 221-8686}
\affiliation{Institut f\"ur Experimentelle Kernphysik, Karlsruher Institut f\"ur Technologie, 76131 Karlsruhe}
\affiliation{Kavli Institute for the Physics and Mathematics of the Universe (WPI), University of Tokyo, Kashiwa 277-8583}
\affiliation{Korea Institute of Science and Technology Information, Daejeon 305-806}
\affiliation{Korea University, Seoul 136-713}
\affiliation{Kyungpook National University, Daegu 702-701}
\affiliation{\'Ecole Polytechnique F\'ed\'erale de Lausanne (EPFL), Lausanne 1015}
\affiliation{Faculty of Mathematics and Physics, University of Ljubljana, 1000 Ljubljana}
\affiliation{University of Maribor, 2000 Maribor}
\affiliation{Max-Planck-Institut f\"ur Physik, 80805 M\"unchen}
\affiliation{School of Physics, University of Melbourne, Victoria 3010}
\affiliation{Moscow Physical Engineering Institute, Moscow 115409}
\affiliation{Graduate School of Science, Nagoya University, Nagoya 464-8602}
\affiliation{Kobayashi-Maskawa Institute, Nagoya University, Nagoya 464-8602}
\affiliation{Nara Women's University, Nara 630-8506}
\affiliation{National Central University, Chung-li 32054}
\affiliation{National United University, Miao Li 36003}
\affiliation{Department of Physics, National Taiwan University, Taipei 10617}
\affiliation{H. Niewodniczanski Institute of Nuclear Physics, Krakow 31-342}
\affiliation{Nippon Dental University, Niigata 951-8580}
\affiliation{Niigata University, Niigata 950-2181}
\affiliation{University of Nova Gorica, 5000 Nova Gorica}
\affiliation{Osaka City University, Osaka 558-8585}
\affiliation{Pacific Northwest National Laboratory, Richland, Washington 99352}
\affiliation{Panjab University, Chandigarh 160014}
\affiliation{University of Pittsburgh, Pittsburgh, Pennsylvania 15260}
\affiliation{University of Science and Technology of China, Hefei 230026}
\affiliation{Seoul National University, Seoul 151-742}
\affiliation{Soongsil University, Seoul 156-743}
\affiliation{Sungkyunkwan University, Suwon 440-746}
\affiliation{School of Physics, University of Sydney, NSW 2006}
\affiliation{Tata Institute of Fundamental Research, Mumbai 400005}
\affiliation{Excellence Cluster Universe, Technische Universit\"at M\"unchen, 85748 Garching}
\affiliation{Toho University, Funabashi 274-8510}
\affiliation{Tohoku Gakuin University, Tagajo 985-8537}
\affiliation{Tohoku University, Sendai 980-8578}
\affiliation{Department of Physics, University of Tokyo, Tokyo 113-0033}
\affiliation{Tokyo Institute of Technology, Tokyo 152-8550}
\affiliation{Tokyo Metropolitan University, Tokyo 192-0397}
\affiliation{Tokyo University of Agriculture and Technology, Tokyo 184-8588}
\affiliation{University of Torino, 10124 Torino}
\affiliation{CNP, Virginia Polytechnic Institute and State University, Blacksburg, Virginia 24061}
\affiliation{Wayne State University, Detroit, Michigan 48202}
\affiliation{Yamagata University, Yamagata 990-8560}
\affiliation{Yonsei University, Seoul 120-749}
 \author{T.~Iwashita}\affiliation{Nara Women's University, Nara 630-8506} 
  \author{K.~Miyabayashi}\affiliation{Nara Women's University, Nara 630-8506} 
  \author{V.~Bhardwaj}\affiliation{Nara Women's University, Nara 630-8506} 
  \author{I.~Adachi}\affiliation{High Energy Accelerator Research Organization (KEK), Tsukuba 305-0801} 
  \author{H.~Aihara}\affiliation{Department of Physics, University of Tokyo, Tokyo 113-0033} 
  \author{D.~M.~Asner}\affiliation{Pacific Northwest National Laboratory, Richland, Washington 99352} 
  \author{T.~Aushev}\affiliation{Institute for Theoretical and Experimental Physics, Moscow 117218} 
  \author{A.~M.~Bakich}\affiliation{School of Physics, University of Sydney, NSW 2006} 
  \author{A.~Bala}\affiliation{Panjab University, Chandigarh 160014} 
  \author{B.~Bhuyan}\affiliation{Indian Institute of Technology Guwahati, Assam 781039} 
% ...
\collaboration{The Belle Collaboration}

\begin{abstract}
We report an observation of the $B^{\pm} \to J/\psi \eta K^{\pm}$ 
and $B^0 \to J/\psi \eta K^0_S$ decays using 
772$\times 10^{6}$ $B\overline{B}$ pairs collected 
at the $\Upsilon(4S)$ resonance with the Belle detector at the KEKB 
asymmetric-energy $e^+e^-$ collider. 
We obtain the branching fractions 
${\cal B}(B^{\pm}\rightarrow J/\psi\eta K^{\pm})=(1.27\pm 0.11{\rm (stat.)\pm 0.11{\rm (syst.)})}\times10^{-4}$ and 
${\cal B}(B^0\to J/\psi \eta K^0_S)=(5.22 \pm 0.78 {\rm(stat.)} \pm 0.49{\rm(syst.)})\times10^{-5}$.
We search for a new narrow charmonium(-like) state $X$ in the 
$J/\psi \eta$ mass spectrum and find no significant excess. 
We set upper limits on the product of branching fractions,
${\cal B}(B^\pm \to XK^\pm){\cal B}(X \to J/\psi \eta)$, at 3872 MeV$/c^2$ 
where a $C$-odd partner of $X(3872)$ may exist, at $\psi(4040)$ and 
$\psi(4160)$ assuming their known mass and width, and over a range from 
3.8 to 4.8 GeV$/c^2$. 
The obtained upper limits at 90\% confidence level for 
$X^{C{\rm -odd}}(3872)$, $\psi(4040)$ and $\psi(4160)$ are 
3.8$\times 10^{-6}$, 15.5$\times 10^{-6}$ and 
\black{7.4}$\times 10^{-6}$, respectively.
\end{abstract}

\maketitle

\tighten

{\renewcommand{\thefootnote}{\fnsymbol{footnote}}}
\setcounter{footnote}{0}

The discovery of a narrow charmonium-like resonance, $X(3872)$, in 
the $J/\psi\pi^+\pi^-$ final state by the Belle collaboration in 
2003~\cite{disx3872} opened a new era in the spectroscopy 
of charmonium and charmonium-like exotic states~\cite{nora}.
In addition to $J/\psi\pi^+\pi^-$, $X(3872)$ decays are also seen in the
$D^0 \bar{D}^{*0}$~\cite{dd}, $J/\psi\pi^+\pi^-\pi^0$~\cite{jpo} and 
$J/\psi\gamma$~\cite{jpg1,jpg3} final states. 
Observation of the $X(3872)\to J/\psi\gamma$ mode confirms that its 
$C$-parity is even. 
The studies of angular distributions of the decay products in the 
$X(3872)\to J/\psi\pi^+\pi^-$ mode by CDF~\cite{bib_cdf_x3872jpc} and 
Belle~\cite{bib_belle_x3872prd} as well as the $3\pi$ invariant mass spectrum 
in  $J/\psi\pi^+\pi^-\pi^0$ mode by BaBar~\cite{jpo} restrict $J^{PC}$ to be 
either $1^{++}$ and $2^{-+}$ but do not allow 
a definitive determination.
A full five-dimensional amplitude analysis of the angles among 
the decay products in $B^+ \to X(3872) K^+,~X(3872) \to J/\psi \pi^+\pi^-$ 
recently performed by the LHCb collaboration has unambiguously
assigned $J^{PC}=1^{++}$ to the $X(3872)$~\cite{bib_lhcb_x3872}.

The very small width ($\Gamma < 1.2$ MeV)~\cite{bib_belle_x3872prd} 
of the $X(3872)$ and its mass ($M=3871.7\pm0.2$ MeV/$c^2$) close to 
the $D^0{\bar D^{*0}}$ threshold~\cite{pdg} make its interpretation
as a $D^0{\bar D^{*0}}$ molecule~\cite{abx3872} quite plausible.
However, other models such as tetraquark~\cite{tetra}, hybrid  
($c{\bar c}g$)~\cite{hybrid} and the admixture of molecular and charmonium 
states~\cite{mixing} are not excluded. 
In both the molecule and tetraquark 
pictures~\cite{bib_terasaki, th_nieves_pavon}, a $C$-odd partner 
($X^{C{\rm -odd}}$) or a charged partner ($X^{\pm}$) of $X(3872)$ can exist. 
So far, searches for the charged partner 
$X^{\pm}\to J/\psi \pi^{\pm}\pi^{0}$ have given negative 
results~\cite{bib_babar_jpsipicpi0, bib_belle_x3872prd}. 
This might be only because the $X^{\pm}$ is too broad, given the 
current statistics; it leaves open the possibility of a 
moderately narrow $C$-odd partner, as postulated by the tetraquark 
model~\cite{bib_terasaki}. Recently, the Belle collaboration has searched 
for the $X^{C{\rm -odd}}\to \chi_{c1}\gamma$ transition in $B\to\chi_{c1}\gamma K$ 
decays and reported evidence for a narrow 
resonance at 3823 MeV/$c^2$~\cite{bib_belle_chic1gam}. 
This resonance is presumably the $1^{3}D_2$ $c\bar{c}$ $(\psi_2)$ 
rather than the $X^{C{\rm -odd}}$, 
since its mass, decay width and the discovery decay mode are consistent 
with theoretical prediction 
for this charmonium state~\cite{fr_vish1,fr_vish2,fr_vish3}.

Alternatively, the $X^{C{\rm -odd}}$ might appear in the $J/\psi \eta$ 
final state.
The photon energy in $\eta \to \gamma \gamma$ is well above 
the energy threshold to be detected in $B$-factory experiments 
even in the case where the resonance is just above 
the $J/\psi \eta$ mass threshold. 
Therefore, the $J/\psi \eta$ system in the three-body 
$B \to J/\psi \eta K$ decay is a suitable final state to search for a 
missing $C$-odd partner of the $X(3872)$ as well as any yet-unseen 
charmonium(-like) resonances.
The $J/\psi\eta$ final state is also sensitive to the $\psi(4040)$ 
and $\psi(4160)$ resonances, whose decays into $J/\psi \eta$ were recently 
reported by BESIII in $e^+e^-$ annihilation~\cite{bib_bes3_4040} and Belle in the initial state radiation 
process~\cite{bib_belle_isr_psieta}.
Since the total width and partial width to $e^+e^-$ are known for 
these charmonia~\cite{pdg}, this observation implies $\psi(4040)$ 
and $\psi(4160)$ have branching fractions of a few percent to $J/\psi \eta$. 
If the branching fractions for $B^{\pm} \to \psi(4040) K^{\pm}$ or 
$\psi(4160) K^{\pm}$ are as high as $\sim 10^{-3}$, these decay channels are 
accessible with Belle's data set.

The branching fraction for $B\to J/\psi \eta K$ decay may also shed 
light on the inclusive spectrum of $B\to J/\psi X$, which 
is fairly well described by non-relativistic QCD
calculations~\cite{bib_nrqcd} except for an excess in the low momentum
region~\cite{bib_incl_cleo,bib_babar_incl}.
There have been several models proposed to explain this excess, such as
$B \to J/\psi K_g$ (where $K_g$ is a hybrid meson with $\bar{s}qg$
constituents)~\cite{bib_sqg}, or a still-undiscovered 
charmonium(-like) state that decays into 
$J/\psi$~\cite{bib_yetxyz}.
Such exotic or new states can be constrained by measurements 
of \black{multibody} $B$ decay modes into $J/\psi$, such as $B\to J/\psi\eta K$, 
because they populate  
the region of the above-mentioned excess.

A previous study by the BaBar collaboration~\cite{babarjek} reported an 
observation of $B^{\pm} \to J/\psi \eta K^{\pm}$ and evidence for 
$B^0 \to J/\psi \eta K^0_S$~\cite{cc} using $90\times10^6$ $B\overline{B}$ 
pairs \black{($N_{B{\bar B}}$)}, but no signal
of a narrow resonance \black{was} found in the $J/\psi \eta$ spectrum. 
In this paper, we present a study of $B\to J/\psi\eta K$ decays based on a 
data sample of 772$\times 10^{6}$ $B\overline{B}$ events collected with the 
Belle detector at the KEKB asymmetric-energy $e^+e^-$ collider~\cite{KEKB} 
at the $\Upsilon(4S)$ resonance. 

The Belle detector is a large solid-angle magnetic spectrometer that 
consists of a silicon vertex detector (SVD), a 50-layer central drift 
chamber (CDC), an array of aerogel threshold Cherenkov counters (ACC), a 
barrel-like arrangement of time-of-flight scintillation counters (TOF), 
and an electromagnetic calorimeter (ECL) comprised of CsI(Tl) crystals 
located inside a superconducting solenoid coil that provides a 
1.5~T magnetic field.
An iron flux return located outside of the coil is instrumented to 
detect $K_L^0$ mesons and to identify muons (KLM).
The detector is described in detail elsewhere~\cite{Belle}.
Two inner detector configurations were used. 
A 2.0 cm radius beampipe and a three-layer silicon vertex detector were used 
to collect the first sample of 152$\times 10^{6}$ 
$B\overline{B}$ pairs, while a 1.5 cm radius beampipe, a four-layer silicon 
detector and a small-cell inner drift chamber were used to record the 
remaining 620$\times 10^{6}$ 
$B\overline{B}$ pairs~\cite{svd2}.

Charged tracks coming from $B$ decays should originate from 
the interaction point (IP). 
The closest approach with respect to the IP is required to be within 
5.0 cm along the beam direction ($z$-axis) and within 2.0 cm in the transverse plane.
Photons are reconstructed as ECL clusters \black{without an} 
associated charged \black{track that have} transverse shower shape variables 
consistent with an electromagnetic cascade hypothesis\black{. 
For $\eta$ reconstruction, the daughter photon} has an energy greater 
than 100 MeV in the laboratory frame.

The $J/\psi$ meson is reconstructed in its decay to 
$\ell^+\ell^-$ ($\ell$ $=$ $e$ or $\mu$).
From the selected charged tracks, $e^{\pm}$ candidates are identified 
by combining specific-ionization ($dE/dx$) information from the CDC, $E/p$ (\black{where} $E$ is 
the shower energy detected in the ECL and $p$ is the momentum measured by 
the SVD and the CDC), and shower shape in the ECL.
In addition, the ACC information and \black{the} position \black{difference} 
between the electron track \black{candidate} and the \black{matching ECL cluster} 
are used in 
the identification of electron candidates. 
In the $J/\psi\to e^+e^-$ mode, in order to 
recover bremsstrahlung photons and final state radiation,
the four-momenta of all photons within 50 mrad of each of the leptons are included in the 
invariant mass that \black{is  hereinafter denoted as $M_{e^+e^-(\gamma)}$.} 
Identification of $\mu$ candidates is based on the track penetration depth 
and hit pattern in the KLM system~\cite{muid}. 
The reconstructed invariant mass of \black{a} $J/\psi$ \black{candidate must} 
satisfy 2.95 GeV$/c^2$ $<$ $M_{e^+ e^-(\gamma)}$ $<$ 3.13 GeV$/c^2$ or 
3.04 GeV$/c^2$ $<$ $M_{\mu^+ \mu^-}$ $<$ 3.13 GeV$/c^2$.  
In order to improve the momentum resolution, 
\black{a vertex-constrained fit followed by a mass-constrained fit 
is applied} for the $J/\psi$ candidates and convergence of \black{both} fits 
is required. 

\black{Pairs of photons are combined to form $\eta$ candidates} 
within \black{the} mass range 510 MeV$/c^2$ $<$ $M_{\gamma\gamma}$ $<$ 
575 MeV$/c^2$.
\black{To further reduce combinatorial background, the $\eta \to \gamma \gamma$ candidates are required to have an energy balance parameter $(|E_1-E_2|/(E_1+E_2))$ smaller than 0.8, where $E_1~(E_2)$ is the energy of the first (second) photon in the laboratory frame.}
\black{To suppress} the background \black{photons from $\pi^0$ decays}, 
we \black{ reject any photon} forming a $\pi^0$ \black{candidate} 
(117 MeV$/c^2$ $<$ $M_{\gamma\gamma}$ $<$ 153 MeV$/c^2$) with any other 
photon in the event.
For the selected $\eta$ candidates, a mass-constrained fit is performed 
to improve the momentum resolution.
 
Charged kaons are identified by combining information from the CDC, TOF and 
ACC systems. 
The kaon identification efficiency is about 90\% while the probability of 
misidentifying a pion as a kaon is about 10\% for the corresponding momentum range. 
$K^0_S$ mesons are reconstructed 
by combining two oppositely charged tracks (both assumed to be pions) and 
requiring the invariant mass $M_{\pi^+\pi^-}$ to be between 
482 and 514 MeV/$c^2$.
The selected candidates are required to have a 
vertex \black{displaced} from the IP as described in Ref.~\cite{ks}.

A $B\to J/\psi\eta K$ candidate is formed from the $J/\psi$, $\eta$ and 
kaon \black{candidates and} is identified by two kinematic variables defined 
in the $\Upsilon(4S)$ rest frame (cms): the energy difference 
$(\Delta E \equiv E^*_B-E^*_{\rm beam})$ and the beam-energy constrained mass 
$(M_{\rm bc} \equiv \sqrt{(E^*_{\rm beam})^2-(P^{*}_B)^2})$.
Here, $E^*_{\rm beam}$ is the cms beam energy and $E^*_B$ and $P^*_B$ are the 
cms energy and momentum, respectively, of the reconstructed $B$ candidate.
Events having at least one $B$ candidate satisfying 
$M_{\rm bc} > 5.27$ GeV/$c^2$ and 
$|\Delta E|<0.2$ GeV are retained for further analysis.

Among \black{the retained} events, \black{29\% have} more than one $B$ candidate. 
This is predominantly due to \black{the} wrong combination \black{in} forming 
the $\eta$ candidate \black{or, far} less frequently\black{,} due to an incorrect 
$J/\psi\to \ell^+\ell^-$ \black{reconstruction; cases} with an incorrect kaon 
candidate \black{are} negligible.
Therefore, we select the $B$ candidate having the smallest 
\black{goodness of fit,} defined as 
$\chi^2 \equiv(M_{\ell^+\ell^-} - m_{J/\psi})^2/\sigma_{\ell^+\ell^-}^2 + (M_{\gamma \gamma} - m_{\eta})^2/\sigma_{\gamma \gamma}^2$, 
where $M_{\ell^+\ell^-}$ denotes $M_{e^+e^-(\gamma)}$ or 
$M_{\mu^+\mu^-}$, $\sigma_{\ell^+\ell^-}$ denotes the 
$M_{\ell^+\ell^-}$ resolutions (11.1 MeV/$c^2$ for $M_{e^+e^-(\gamma)}$ and 
8.9 MeV/$c^2$ for $M_{\mu^+\mu^-}$), $M_{\gamma \gamma}$ is the photon pair 
mass, and $\sigma_{\gamma\gamma}$ is the $M_{\gamma\gamma}$ resolution 
(13.8 MeV$/c^2$). 
Here, $m_{J/\psi}$ and $m_{\eta}$ are \black{the} nominal \black{meson} 
masses~\cite{pdg}.

To suppress continuum background, we  reject events having a ratio 
\black{$R_2$} of the second to zeroth Fox-Wolfram moments~\cite{r2} greater 
than 0.5.
Among the backgrounds from $B\overline{B}$ events\black{,} those that contain a 
real $J/\psi \to \ell^+\ell^-$ decay dominate.
\black{A large sample of} $B\to J/\psi X$ Monte Carlo (MC) decays\black{,} 
corresponding to 100 times the data sample\black{,} is used to model \black{this} 
background \black{component's} $M_{\rm bc}$ and $\Delta E$ distributions.
When $\psi^{\prime}$ decays to the final states other than $J/\psi\eta$, the 
$B \to \psi^{\prime} K$ decay mode \black{forms a} significant portion of 
\black{the} background.
\black{We} denote this contribution as the 
$B\to\psi^{\prime}(\not\to J/\psi\eta)K$ process.
In order to reduce this background, we reject a $J/\psi$ 
\black{that, when combined with a $\pi^+\pi^-$ pair,} forms a $\psi^{\prime}$ 
candidate \black{with a} mass difference \black{in the} range 
0.58 GeV$/c^2$ $< M_{J/\psi\pi^+\pi^-}-m_{J/\psi} <$ 0.60 GeV$/c^2$.
The non-$J/\psi$  background is estimated using the $M_{\ell^+\ell^-}$ 
sideband events in data and \black{is} found to be negligible.

\begin{figure}[htb!]
\includegraphics[width=0.47\textwidth,clip]{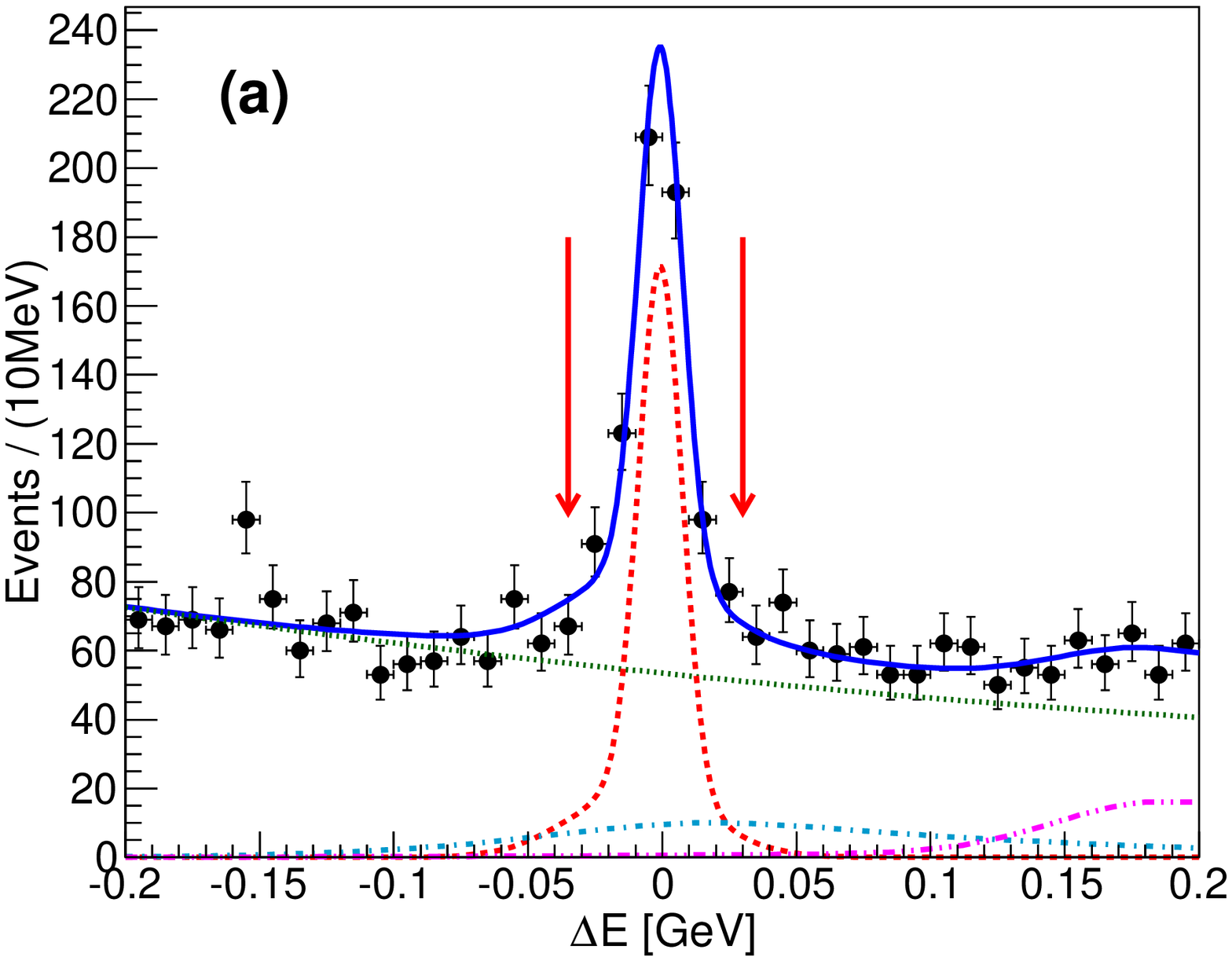}
\includegraphics[width=0.47\textwidth,clip]{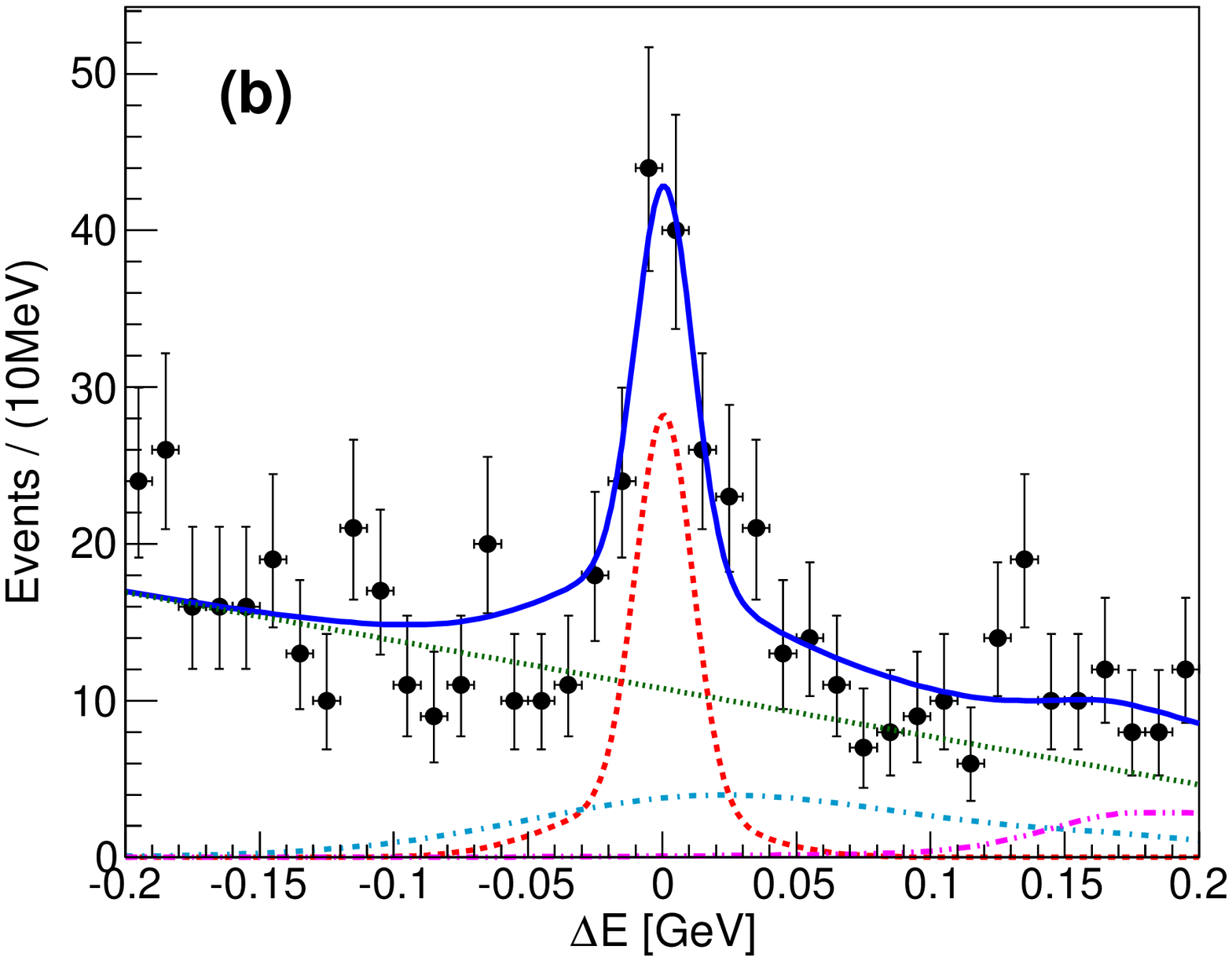}
\caption{(color online). $\Delta E$ distribution of (a) 
$B^{\pm}\to J/\psi\eta K^{\pm}$ and (b) $B^0\to J/\psi\eta K^0_S$ candidates 
in $M_{\rm bc}$ $>$ 5.27 GeV/$c^2$. Signal-enhanced region for 
$B^{\pm}\to J/\psi \eta K^{\pm}$ is shown by the the red arrows in (a). 
Data are shown by points with error bars. 
The red dashed line is signal, the cyan dot-dashed line is 
$B\to \psi^{\prime}(\not\to J/\psi\eta)K$ background, the magenta 
dot-dot-dashed 
line is $B\to\chi_{c1}K$ background and the green dotted line is 
other backgrounds.}
\label{fig_deltae}
\end{figure}

The $B$ decay signal extraction is carried out by performing an extended 
unbinned maximum likelihood (UML) fit to the $\Delta E$ distribution.
Figure.~\ref{fig_deltae} shows the $\Delta E$ distribution 
for the charged and neutral $B$ decay candidates 
together with the fit results. 
\black{Clear signal peaks are seen on smoothly distributing background for both cases.}
For these decays, a sum of two Gaussians is used to model the probability 
density function (PDF) for signal events.
For the $B^{\pm}\to J/\psi\eta K^{\pm}$ decay mode, the mean and width of 
the core Gaussian are floated and the remaining parameters are fixed 
\black{to values obtained} by fitting the signal MC \black{distribution}. 
Since we have smaller statistics for $B^0 \to J/\psi \eta K^0_S$, 
\black{the parameters of the signal PDF is fixed to the values of data obtained by }
the $B^{\pm}\to J/\psi\eta K^{\pm}$ sample. 
Since the $B\to\psi^{\prime}(\not\to J/\psi\eta)K$ and $B\to\chi_{c1}K$ 
decay modes are expected to have different features compared to other 
backgrounds in the $\Delta E$ distribution, these two processes are treated 
separately.
We use a bifurcated Gaussian to describe these 
decay modes whose parameters are fixed from large MC simulation samples.
Since the branching fractions for \black{these} decay modes 
are known~\cite{pdg}, their yields are also fixed.
To model the remaining featureless combinatorial background in the $\Delta E$ 
projection, we use \black{a second-order (first-order)} Chebyshev polynomial 
for the $B^{\pm} \to J/\psi \eta K^{\pm}$ ($B^0\to J/\psi\eta K^0_S$) decay 
mode.
We obtain signal yields of $428\pm37$ events and $94\pm14$ events for 
the $B^{\pm}\to J/\psi\eta K^{\pm}$ and $B^0\to J/\psi\eta K^0_S$ 
decay modes, respectively. 
The detection efficiency estimation for $B^{\pm} \to J/\psi \eta K^{\pm}$ is described in more detail later. The three-body phase space
distribution is assumed for $B^0 \to J/\psi \eta K^0_S$.
Their branching fractions are $\black{(1.27\pm0.11\pm0.11)}\times10^{-4}$ 
and $(5.22\pm0.78\pm0.49)\times10^{-5}$, where the first uncertainty is statistical and the second is systematic uncertainty; these uncertainties are described later in detail.
We calculate the statistical significance,
$\sqrt{-2\ln{\mathcal{L}_0/\mathcal{L}_{\rm max}}}$, where 
$\mathcal{L}_{\rm max}$ ($\mathcal{L}_0$) denote the 
likelihood value when the signal yield is allowed to vary (is set to zero). 
The significance 
is found to be 17$\sigma$ (7$\sigma$) for the 
$B^{\pm} \to J/\psi \eta K^{\pm}$ ($B^0 \to J/\psi \eta K^0_S$) decay mode.
\black{We observe the $B^0 \to J/\psi \eta K^0_S$ decay mode for the first time with the significance more than 5 $\sigma$.}
Equal production of neutral and charged $B$ meson \black{pairs 
in} the $\Upsilon(4S)$ decay is assumed. We used the secondary 
branching fractions \black{reported} in Ref.~\cite{pdg}. 
\black{The results of the fits are presented in Table~\ref{tb_brall}.}
 
\begin{table*}[htb!]
\caption{\black{Summary of the detection efficiency ($\epsilon$), 
signal yield ($N_{\rm sig}$) and branching fraction (${\cal B}$) 
in $-0.2$ GeV$/c^2$ $<$ $\Delta E$ $<$ $0.2$ GeV$/c^2$, where the first and 
second errors are statistical and systematic.}}
\label{tb_brall}
\begin{tabular}
{@{\hspace{0.5cm}}l@{\hspace{0.5cm}}  @{\hspace{0.5cm}}c@{\hspace{0.5cm}} @{\hspace{0.5cm}}c@{\hspace{0.5cm}} @{\hspace{0.5cm}}c@{\hspace{0.5cm}} @{\hspace{0.5cm}}c@{\hspace{0.5cm}}}
\hline \hline
Decay mode & $\epsilon$(\%) & $N_{\rm sig}$ & ${\cal B}$ \\
\hline
$B^{\pm}\to J/\psi\eta K^{\pm}$ & 9.37 & 428$\pm$37 & \black{(1.27$\pm$0.11$\pm$0.11)}$\times10^{-4}$\\
$B^0\to J/\psi\eta K^{0}_S$ & 7.23 & 94$\pm$14 & (5.22$\pm$0.78$\pm$0.49)$\times10^{-5}$\\
\hline \hline
\end{tabular}
\end{table*}

\black{Since the $B^{\pm} \to J/\psi \eta K^{\pm}$} 
signal is strong, we use the $J/\psi \eta$ mass \black{spectrum} ($M_{J/\psi\eta}$) to 
resolve the intermediate states in this three-body final state.
For this purpose, we select events having $-35$ MeV $<$ $\Delta E$ $<$ $30$ MeV. 
This requirement \black{corresponds to $\pm3.5\sigma$ ($\pm1.3\sigma$)} of the 
narrower (wider) Gaussian.
The $B$ decay signal yield in this signal-enhanced region is 
$403\pm35$ events.

\begin{figure}[htb!]
\includegraphics[width=0.48\textwidth,clip]{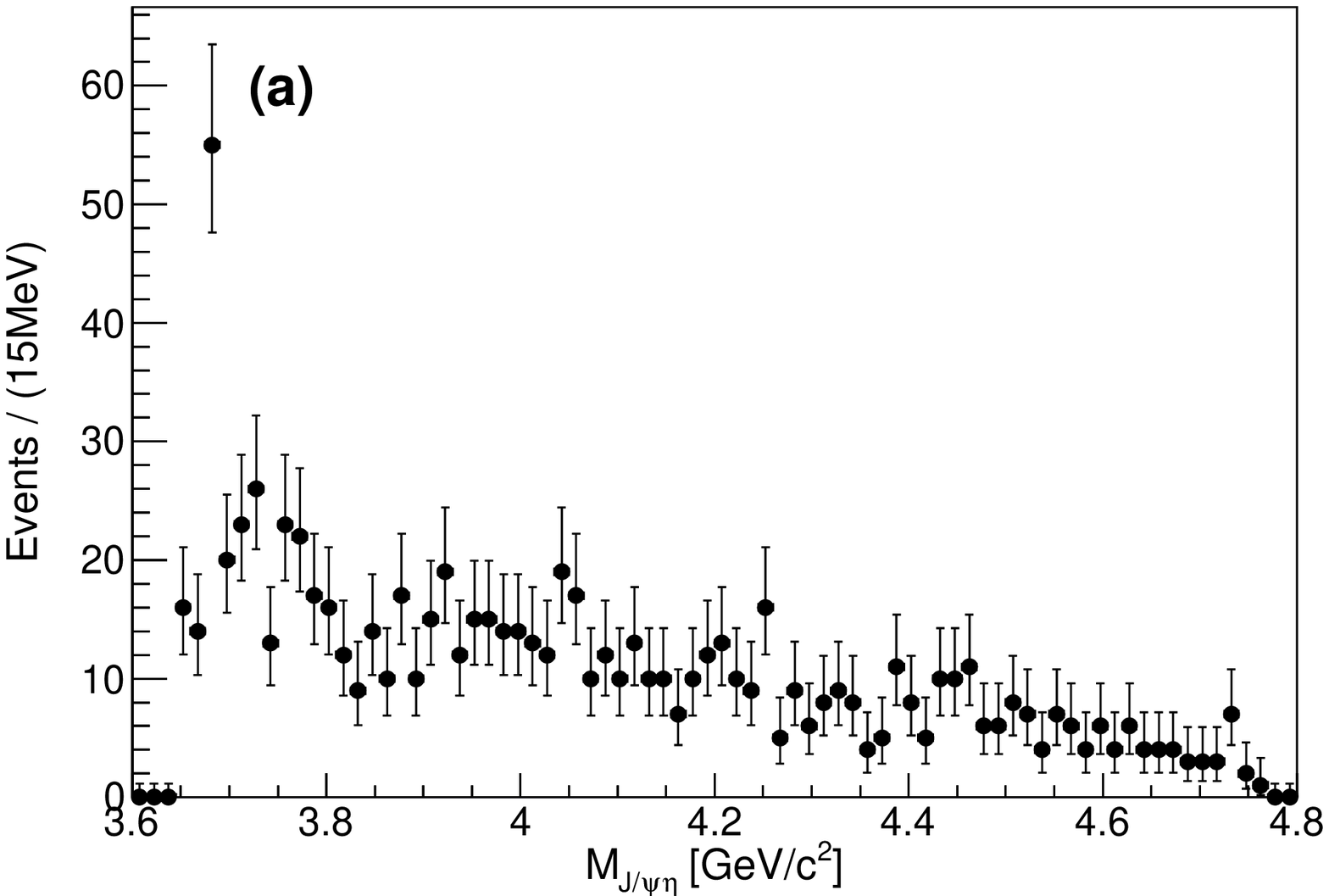}
\includegraphics[width=0.48\textwidth,clip]{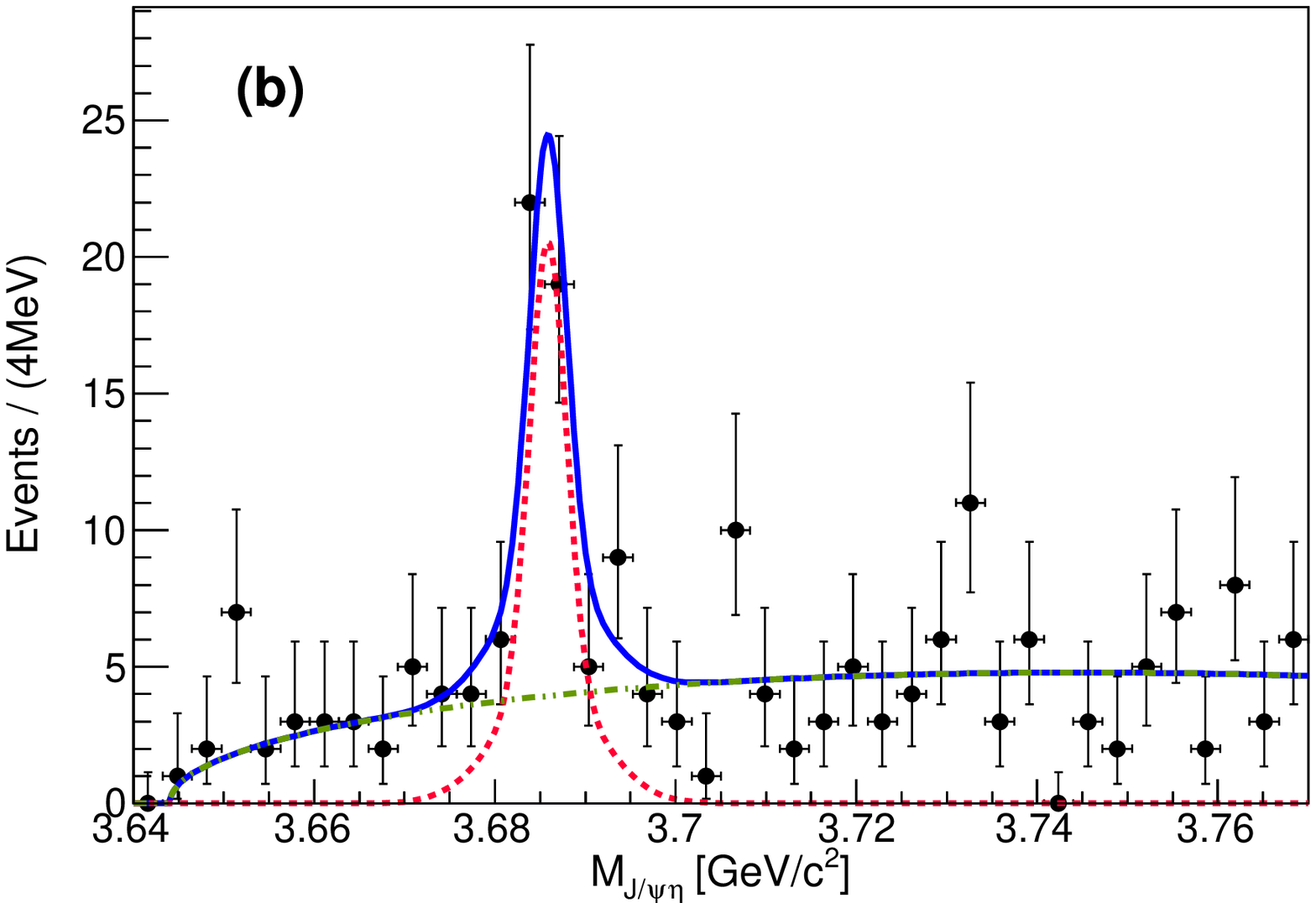}
\includegraphics[width=0.48\textwidth,clip]{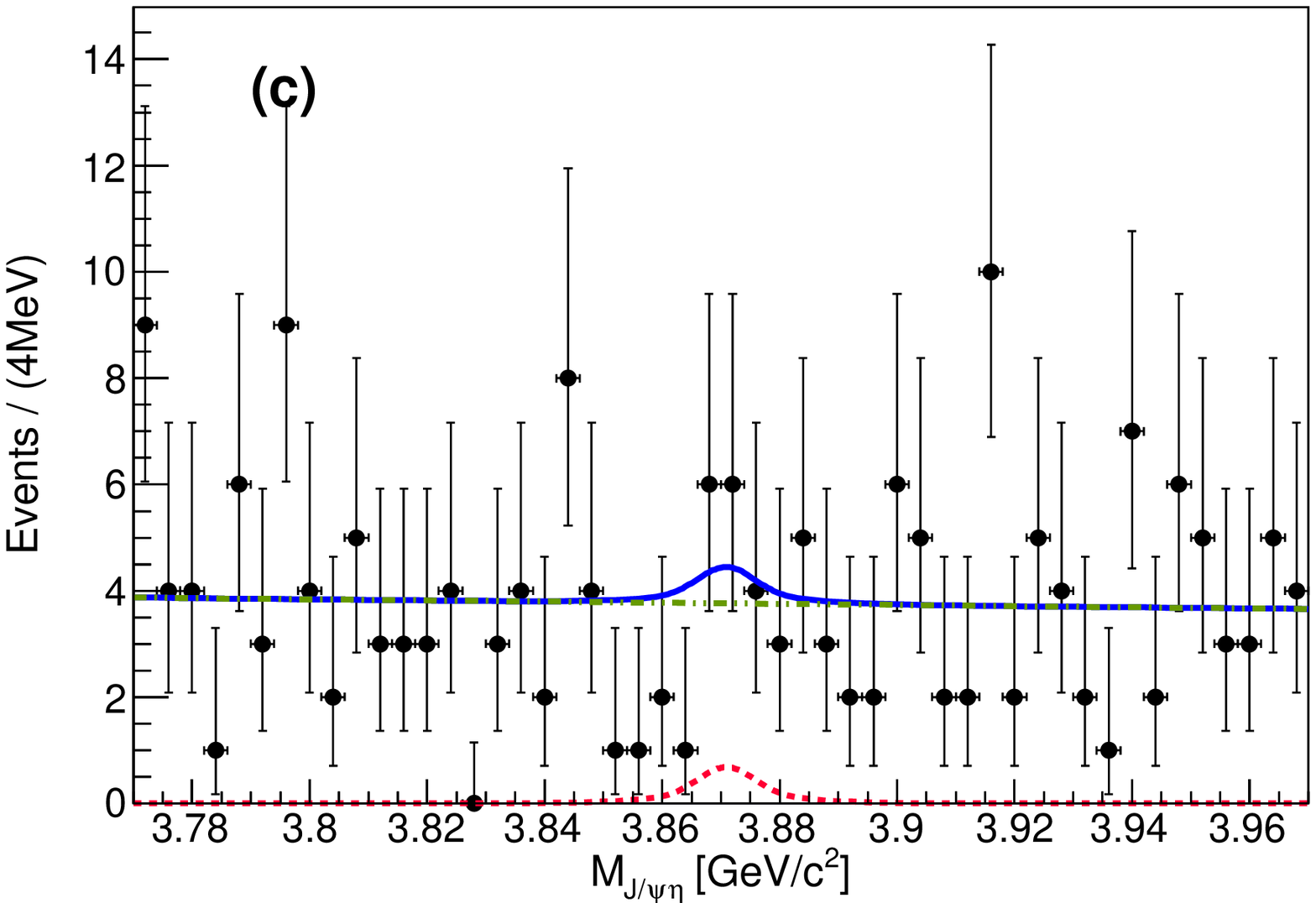}
\caption{(color online). \black{The $J/\psi\eta$ invariant mass ($M_{J/\psi \eta}$) distribution of $B^{\pm} \to J/\psi \eta K^{\pm}$ candidates for: 
(a) the entire mass distribution, (b) the region around the $\psi^{\prime}$ and (c) the $X(3872)$ region. Data is shown by points with error bars; overall fit is shown by blue solid line. 
For (b) and (c), the red dashed line is for signal 
($\psi'$ and $X(3872)$ in (b) and (c), respectively) and the green two 
dotted-dashed line is for the remainder.}
}
\label{fig_mjeta}
\end{figure}

The $M_{J/\psi \eta}$ distribution for this \black{subsample} is shown in 
Fig.~\ref{fig_mjeta}(a).
We find a clear peak corresponding to the $\psi^{\prime} \to J/\psi \eta$ decay
at 3686 MeV/$c^2$ \black{with a yield of} 46$\pm$8 events by performing 
an UML fit to the $M_{J/\psi \eta}$ distribution in the range from \black{the} 
kinematical threshold to 3770 MeV/$c^2$. We \black{parametrize} the 
$\psi^{\prime}$ signal and \black{remaining} contributions by the sum of two 
\black{Gaussians and a threshold function, respectively, as shown in Fig.~\ref{fig_mjeta}(b).} The $\psi^{\prime}$ shape is fixed to 
that found by a fit to the MC \black{distribution}, which is 
\black{calibrated by} the 
difference in resolution between data and simulation. 
The \black{$M_{J/\psi\eta}$ calibration factor} is taken from the $\Delta E$ 
distribution, \black{since both} resolution\black{s are} dominated by that 
\black{of} the $\eta$ \black{(}reconstructed from photons rather than charged 
tracks\black{)}.
The threshold function is taken as 
$a(M_{J/\psi\eta}-m_0)^{1/2}+b(M_{J/\psi\eta}-m_0)^{3/2}+c(M_{J/\psi\eta}-m_0)^{5/2}$, 
where $m_0$ $=$ 3.644 GeV/$c^2$ and the shape determined by 
$a$, $b$ and $c$ is fixed to MC simulation\black{;} its normalization is 
floated in the fit. 
\black{We obtain ${\cal B}(B^{\pm}\to \psi^{\prime}K^{\pm}){\cal B}(\psi^{\prime}\to J/\psi \eta)=(0.15\pm0.03{\rm (stat.)}\pm0.01{\rm (syst.)})\times 10^{-4}$, which is in agreement with the PDG value~\cite{pdg}.}
The rest of the $B$ decay signal does not show any peaking structure and 
is consistent with three-body phase space. 

\begin{figure}[htb!]
\includegraphics[width=0.6\textwidth,clip]{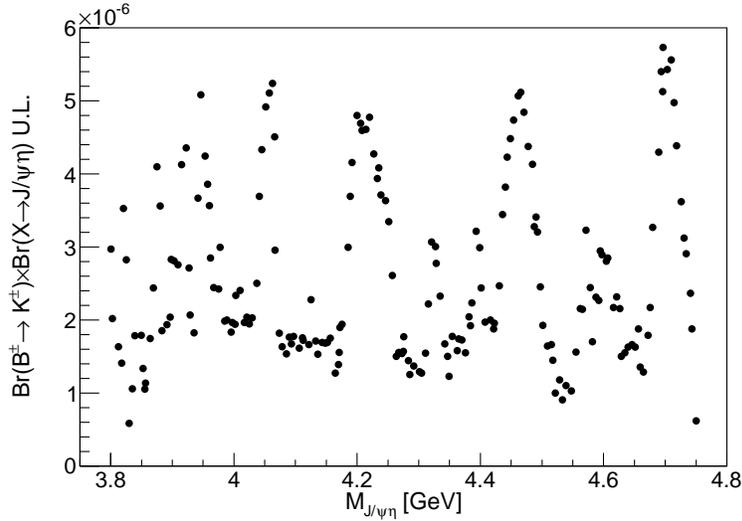}
\caption{90\% C.L. upper limit of the ${\cal B}(B^{\pm}\to XK^{\pm}){\cal B}(X\to J/\psi\eta)$ \black{for a narrow resonance $X$ as a} function of the mass, with a 5 MeV/$c^2$ interval.}
\label{fig_ul}
\end{figure}

\black{The efficiency that is used to obtain the total branching fraction is determined by weighting the $B^{\pm}\to\psi^{\prime}K^{\pm}$ and the three-body phase components according to the observed $M_{J/\psi\eta}$ spectrum.}
\black{After subtracting the yield of 46$\pm$8 events for} 
$B^{\pm}\to\psi^{\prime} K^{\pm}$ followed 
by $\psi^{\prime}\to J/\psi\eta$ \black{(as described earlier), the remaining} 
$B$ decay signal yield \black{is} 357$\pm$38 events \black{and is used to 
extract the} branching fraction \black{in Table~\ref{branch}}. 

\begin{table*}[htb!]
\caption{Summary of the detection efficiency ($\epsilon$), signal 
yield ($N_{\rm sig}$) and branching fraction (${\cal B}$), where the first 
and second errors are statistical and systematic, respectively. 
\black{For $B^{\pm}\to \psi^{\prime}  K^{\pm}$, followed by 
$\psi^{\prime}\to J/\psi\eta$, ${\cal B}$ denotes the products of the 
branching fractions,} 
${\cal B}(B^{\pm}\to \psi^{\prime}  K^{\pm}){\cal B}(\psi^{\prime}\to J/\psi\eta)$. For the $B^{\pm}$ decays, 
all relevant numbers are defined in the signal enhanced region, 
$-35$ MeV $<$ $\Delta E$ $<$ 30 MeV.}
\label{branch}
\begin{tabular}
{@{\hspace{0.5cm}}l@{\hspace{0.5cm}}  @{\hspace{0.5cm}}c@{\hspace{0.5cm}} @{\hspace{0.5cm}}c@{\hspace{0.5cm}} @{\hspace{0.5cm}}c@{\hspace{0.5cm}} @{\hspace{0.5cm}}c@{\hspace{0.5cm}}}
\hline \hline
Decay mode & $\epsilon$(\%) & $N_{\rm sig}$ & ${\cal B}$ \\
\hline
$B^{\pm}\to J/\psi\eta K^{\pm}$ (Total) & \black{8.82} & 403$\pm$35 & \black{(1.27$\pm$0.11$\pm$0.11)}$\times10^{-4}$\\
$B^{\pm}\to \psi^{\prime}  K^{\pm}$, $\psi^{\prime}\to J/\psi\eta$ & \black{8.42} & 46$\pm$8 & \black{(0.15$\pm$0.03$\pm$0.01)}$\times10^{-4}$\\
$B^{\pm}\to J/\psi\eta K^{\pm}$ (excl. $\psi^{\prime} K^{\pm}$) & \black{8.88} & 357$\pm$38 & \black{(1.12$\pm$0.11$\pm$0.10)}$\times10^{-4}$\\
\hline \hline
\end{tabular}
\end{table*}

The major source of systematic uncertainty in the branching fraction 
measurements is from the PDF uncertainty. 
\black{
It is estimated by varying all fixed parameters by $\pm 1\sigma$ }
and summing all the variations in quadrature; it amounts to 
7.3\% for $B^{\pm} \to J/\psi \eta K^{\pm}$ and 
8.4\% for $B^0 \to J/\psi \eta K^0_S$. 
The uncertainty of the tracking efficiency is estimated 
to be 0.35\% per track. 
Small differences in the lepton and kaon identification efficiency 
between the data and MC simulation are included in the 
detection efficiency estimation and \black{the} relevant uncertainty is assigned 
as a systematic error. The uncertainty of electron identification is 
studied using the $J/\psi \to e^+e^-$ sample and estimated to be 
0.9\% per $e^+e^-$ pair.
A similar approach for muon identification results in a systematic error 
of 3.9\% per $\mu^+\mu^-$ pair.
A kaon identification uncertainty is determined to be 1.4\% from the study 
using the $D^{*+} \to D^0 (\to K^-\pi^+) \pi^+$ sample.
The uncertainty on the $\eta\to\gamma\gamma$ efficiency is estimated 
to be 3.0\%~\cite{effeta}.
The $K^0_S$ efficiency contributes a 0.7\% error in the 
$B^0 \to J/\psi \eta K^0_S$ mode.
The uncertainties due to signal MC simulation statistics 
(0.5\%) and the secondary branching fractions (0.7\%) have only a small 
effect. The uncertainty of $N_{B \overline{B}}$ 
is 1.4\%. 
Table~\ref{tb_syst} summarizes the systematic uncertainties. 
The overall systematic error is obtained by adding 
all the contributions in quadrature\black{; it is} \black{8.6\%} for 
$B^{\pm} \to J/\psi \eta K^{\pm}$ and \black{9.4\%} for $B^0 \to J/\psi \eta K^0_S$. 

\begin{table}[htb]
\caption{Contributions to the systematic uncertainty of the branching 
fraction. \black{The value of PDF shape in parenthesis is for the $B^{\pm}\to \psi^{\prime}K^{\pm}$ decay followed by $\psi^{\prime}\to J/\psi\eta$.}}
\label{tb_syst}
\begin{tabular}{c  c  c}
\hline \hline
Source & \multicolumn{2}{c}{Contribution (\%)} \\
       & $B^{\pm} \to J/\psi \eta K^{\pm}$ & $B^0 \to J/\psi \eta K^0_S$ \\
\hline
PDF shape ($B^{\pm}\to\psi^{\prime}K^{\pm}$, $\psi^{\prime}\to J/\psi\eta$) & 7.3 (5.8) & 8.4 \\
Tracking efficiency & 1.05 & 0.7 \\
Lepton identification & 2.4 & 2.4 \\
Charged kaon identification & 1.4 & - \\
$\eta \to \gamma \gamma$ efficiency & 3.0 & 3.0 \\
$K^0_S \to \pi^+\pi^-$ efficiency & - & 0.7 \\
Signal MC simulation stat. & 0.5 & 0.5 \\
Secondary ${\cal B}$ & 0.7 & 0.7 \\
$N_{B\overline{B}}$  & 1.4 & 1.4 \\
\hline
Total (inc. $\psi^{\prime}K^{\pm}$) & 8.6 (7.4) & 9.4 \\
\hline \hline
\end{tabular}
\end{table}

In order to \black{probe the} contribution of the $X^{C{\rm -odd}}$ \black{partner} assuming that it has same mass and width as the $X(3872)$, a sum of two Gaussians for signal and a first-order polynomial for background is used. 
For signal\black{,} all the parameters are fixed after applying the \black{same MC-data shape-parameter calibrations used} in the $\psi^{\prime}$ case. 
The $X(3872)$ region is shown in Fig.~\ref{fig_mjeta}(c).
The fit result for the $X^{C{\rm -odd}}$ yield is found to be 2.3$\pm$5.2 
events and we determine a 90\% confidence level (C.L.) upper limit (U.L.) 
on the product of the branching fractions, 
${\cal B}(B^{\pm} \to X^{C{\rm -odd}} K^{\pm}){\cal B}(X^{C{\rm -odd}}\to J/\psi\eta)<$ 3.8$\times10^{-6}$, using a frequentist approach.
For a given signal yield, a large number of MC simulation 
sets, including signal and background components, are generated according to 
their PDFs, and \black{a fit is} performed to \black{each set}.
The C.L. is determined from the fraction of \black{sets} that give a yield 
larger than the one observed in data. 
The input signal yield is varied until \black{we obtain} 90\% C.L.; this input 
yield is the U.L. for the observed signal yield. To take into account the 
systematic uncertainty, the input signal yield for the simulated sets follows 
a Gaussian distribution \black{whose width corresponds} to the systematic 
uncertainty. This ensures that the yield fluctuations within the simulated 
sets exceeds those due solely to Poisson statistics.
We divide the 3.8 to 4.8 GeV$/c^2$ region 
into five 200 MeV$/c^2$-wide interval 
and use the PDF and efficiency estimated at 4070, 4270, 4470 and 
4670 MeV$/c^2$. 
\black{For the $\psi(4040)$ and $\psi(4160)$ cases, 
we describe the resonance} \black{by a Breit}-Wigner \black{function} with 
the mass and width fixed to the values reported in Ref.\cite{pdg}.
Table~\ref{tab_ul} summarizes the U.L. for the $X^{C{\rm -odd}}$ and $\psi(4040,4160)$.
\black{As shown in Fig.~\ref{fig_ul}, we also provide the U.L. at 90\% C.L. of narrow resonances over a range from 
3.8 to 4.8 GeV$/c^2$, with 5 MeV$/c^2$ steps, using the same procedure as 
for the $X^{C{\rm -odd}}$ U.L. estimation.}

\begin{table}[htb!]
\caption{The U.L. for the product of the branching fractions 
\black{${\cal B}(B^{\pm}\to X(\to J/\psi\eta) K^{\pm})$ $\equiv{\cal B}(B^{\pm}\to XK^{\pm}){\cal B}(X\to J/\psi\eta)$ 
at 3872 and} the $\psi$ states recently found to decay into $J/\psi \eta$. 
Note that $\epsilon$ is the corrected detection efficiency and the signal yield
$N_{\rm sig}$ is given as an U.L. at 90\% confidence level.}
\label{tab_ul}
\begin{tabular}{c c c c}
\hline \hline
$M_X$ or $\psi$ & $\epsilon$(\%) & $N_{\rm sig}$ &\quad ${\cal B}(B^{\pm} \to X(\to J/\psi \eta) K^{\pm})$ \\
\hline
3872 & 8.1 & $<10.6$ & $< 3.8\times 10^{-6}$ \\
$\psi(4040)$ & 9.2 & $< 51.4$ & $< 1.55\times 10^{-5}$ \\
$\psi(4160)$ & \black{9.2} & $< 24.3$ & $< 0.74\times 10^{-5}$\\
\hline \hline
\end{tabular}
\end{table}

In summary, we observe the $B^{\pm}\to J/\psi\eta K^{\pm}$ and 
$B^0\to J/\psi\eta K^0_S$ decay modes and present the most precise 
measurements to date of the branching fractions, \black{${\cal B}(B^{\pm}\to J/\psi\eta K^{\pm}) = \black{(1.27 \pm 0.11{\rm (stat.)}} \pm 0.11{\rm (syst.)})\times 10^{-4}$} and ${\cal B}(B^0\to J/\psi\eta K^0_S) = (5.22 \pm 0.78{\rm (stat.)}\pm \textcolor{black}{0.49}(\mbox{syst.}))\times 10^{-5}$.
For the $B^{\pm} \to J/\psi \eta K^{\pm}$ signal, the $M_{J/\psi \eta}$ 
distribution is used to resolve each possible contribution to search for a 
resonance in the $J/\psi \eta$ final state. Except for the known 
$\psi^{\prime} \to J/\psi \eta$ decay, the $M_{J/\psi \eta}$ spectrum is found to be 
featureless and follows a non-resonant distribution.
Because no signal is seen, we obtain an U.L. on the product of the branching 
fractions, ${\cal B}(B^{\pm}\to X^{C{\rm -odd}}K^{\pm}){\cal B}(X^{C{\rm -odd}}\to J/\psi\eta)$ $<$ $3.8\times10^{-6}$ at 90\% C.L.; this is less than one half of the 
corresponding value \black{in $X(3872) \to J/\psi \pi^+\pi^-$~\cite{pdg}.
While} $\psi(4040)$ and $\psi(4160)$ decays into 
$J/\psi \eta$ are observed in the initial state radiation 
process~\cite{bib_belle_isr_psieta}, production of those charmonia and their
decays to the $J/\psi \eta$ final state in $B$ decays are found to 
be insignificant. 
The obtained U.L.s exclude a large branching fraction, 
${\cal O}(10^{-3}),$ for $B^{\pm} \to \psi(4040) K^{\pm}$ and 
$B^{\pm} \to \psi(4160) K^{\pm}$. 
\black{Nevertheless,} values comparable to $B^{\pm} \to \psi^{\prime} K^{\pm}$ 
or $B^{\pm} \to \psi(3770) K^{\pm}$, ${\cal O}(10^{-4})$, are still possible.
Our results show that either the production of the $C$-odd partner of the 
$X(3872)$ resonance in two-body $B$ decay and/or its decay into $J/\psi \eta$ 
is suppressed.

We thank the KEKB group for excellent operation of the accelerator; the KEK cryogenics group for efficient solenoid operations; and the KEK computer group, the NII, and PNNL/EMSL for valuable computing and SINET4 network support.
We acknowledge support from MEXT, JSPS and Nagoya's TLPRC (Japan); ARC and DIISR (Australia); FWF (Austria); NSFC (China); MSMT (Czechia); CZF, DFG, and VS (Germany); DST (India); INFN (Italy); MEST, NRF, GSDC of KISTI, and WCU (Korea); MNiSW and NCN (Poland); MES and RFAAE (Russia); ARRS (Slovenia); IKERBASQUE and UPV/EHU (Spain); SNSF (Switzerland); NSC and MOE (Taiwan); and DOE and NSF (USA).
This work is partly supported by Grant-in-Aid from MEXT for Scientific Research on Innovative Areas (``Elucidation of New hadrons with a Variety of Flavors''). 



\begin{thebibliography}{99}
\bibitem{cc} Inclusion of charge-conjugate modes is implied.
\end{thebibliography}
\end{document}